\documentclass[superscriptaddress,twocolumn,showpacs,preprintnumbers,amsmath,amssymb,prb]{revtex4}

\usepackage{graphicx}
\usepackage{dcolumn}
\usepackage{bm}

\begin{document}

\preprint{}

\title{High-field phase diagram of the Haldane-gap antiferromagnet
Ni(C$_{5}$H$_{14}$N$_{2}$)$_{2}$N$_{3}$(PF$_{6}$)}

\author{H. Tsujii}
\affiliation{Department of Physics, University of Florida, PO Box
118440, Gainesville, Florida 32611-8440}
\affiliation{RIKEN, Wako, Saitama 351-0198, Japan}
\author{Z. Honda}
\affiliation{Faculty of Engineering, Saitama University,
Saitama 338-8570, Japan}
\author{B. Andraka}
\affiliation{Department of Physics, University of Florida, PO Box
118440, Gainesville, Florida 32611-8440}
\author{K. Katsumata}
\affiliation{RIKEN Harima Institute, Mikazuki, Sayo, Hyogo
679-5148, Japan}
\author{Y. Takano}
\affiliation{Department of Physics, University of Florida, PO Box
118440, Gainesville, Florida 32611-8440}

\date{\today}

\begin{abstract}
We have determined the magnetic phase diagram of the quasi-one-dimensional $S =$ 1
Heisenberg antiferromagnet
Ni(C$_{5}$H$_{14}$N$_{2}$)$_{2}$N$_{3}$(PF$_{6}$) by specific heat measurements
to 150 mK in temperature and 32 T in the magnetic field.
When the field is applied along the spin-chain direction, a new phase
appears at $H_{c2}\approx 14$~T.
For the previously known phases of field-induced order, an accurate determination is made
of the power-law exponents of the ordering temperature near the zero-temperature critical field $H_c$,
owing to the four-fold improvement of the minimum temperature over the previous work.
The results are compared with the predictions based on the Bose-Einstein condensation of triplet
excitations.
Substituting deuterium for hydrogen is found to slightly reduce the interchain exchange.
\end{abstract}

\maketitle

\section{\label{sec:1}Introduction}

One-dimensional (1D) integer-spin Heisenberg antiferromagnets (HAF) are well-known for the
Haldane energy gap \cite{Haldane} between the singlet spin-liquid
ground state and the lowest excited state, which is an $S=1$ triplet.
The application of a magnetic field leads to Zeeman splitting of the
triplet and eventual vanishing of the gap $\Delta$ at
$H_c \sim \Delta/{\rm g}\mu_{\rm B}$, where the energy of the lowest branch
of the split triplet reaches the ground-state level.
At this critical field, the expected quantum transition is
to the Tomonaga-Luttinger spin liquid, \cite{Takahashi,Sachdev} in which
spin correlations decay with characteristic power laws.
This scenario remains robust in an array of 1D integer-spin HAF chains
against the introduction of interchain exchange.
Such a coupling reduces $H_c$ but, provided it is small,
does not destroy the singlet ground state below $H_c$.
Above $H_c$, it leads to finite-temperature long-range order (LRO), which can be
described as the three-dimensional Bose-Einstein condensation
of the lower-branch triplets. \cite{Matsubara,Affleck,Affleck91,Tsvelik,Nikuni}
The nature of the ordered state, including the robustness of the
Tomonaga-Luttinger spin liquid, is of strong current interest.

The first experimental evidence \cite{Buyers} of a Haldane gap was found in CsNiCl$_3$.
This material has a relatively large interchain coupling: $J'/J=0.017$,
where $J'$ and $J$ are interchain and in-chain exchanges, respectively.
As a consequence, N$\acute{\rm e}$el ordering occurs at 4.85 K.
Ni(C$_{2}$H$_{8}$N$_{2}$)$_{2}$NO$_{2}$(ClO$_{4}$) (NENP), with the small $|J'|/J$ ratio
of $4\times 10^{-4}$, was the first Haldane-gap antiferromagnet
with no N$\acute{\rm e}$el ordering. \cite{Renard,Katsumata}
Magnetic susceptibility shows no anomaly indicative of ordering, at least down to 4~mK. \cite{Avenel}
Naturally, this material was a promising candidate for the field-induced LRO at
fields above $H_c$.
However, it has been found that such an order is preempted by the staggered effective field
arising from the staggered {\sl g} tensors of the Ni$^{2+}$ ions within each spin chain.

To date, another spin-1 chain material Ni(C$_{5}$H$_{14}$N$_{2}$)$_{2}$N$_{3}$(PF$_{6}$)
(NDMAP) remains the only laboratory model of a 1D HAF array in which the nature of
the field-induced LRO states has been revealed by experiment. \cite{Honda98}
This distinction owes to the unusually low $H_c$, which is readily accessible to
many experimental probes, as well as the absence of a staggered field.
NDMAP has an orthorhombic structure with the lattice parameters
$a=18.046$~\AA, $b=8.7050$~\AA, and $c=6.139$~\AA, with the
antiferromagnetic spin chains running along the $c$ axis. \cite{Monfort}
According to inelastic neutron scattering, \cite{Zhel01} the in-chain exchange is
$J=33.1$~K, and the relative strengths of the interchain exchanges are
$J'_{b}/J\approx 6\times 10^{-4}$ and $|J'_{a}|/J<10^{-4}$ along the
$b$ and $a$ axes, respectively.
Easy-plane crystal-field anisotropy $DS_{z}^{2}$, where $z$ is the
crystallographic $c$ axis and $D/J=0.25$, \cite{Zhel01} is responsible for the strong
anisotropy of both the gap $\Delta$ and the magnetic phase diagram of
the field-induced LRO states, as has been measured by magnetic
susceptibility, \cite{Honda98} specific heat,
\cite{Honda98,Honda01,Chen} magnetization, \cite{Honda00M,Honda01M}
ESR, \cite{Honda99,Hagiwara} and neutron scattering. \cite{Zhel01,Chen,Zhel02,Zhel03,Zhel04}

The gap energies of the triplet, whose degeneracy is lifted by the
crystal-field anisotropy, have been measured by inelastic neutron
scattering \cite{Zhel01} to be $\Delta_{a}=0.42$ meV,
$\Delta_{b}=0.52$ meV, and $\Delta_{c}=1.9$ meV for the $S_{\tilde {z}}=0$
excitations quantized along the $a$, $b$, and $c$ axes, respectively.
Here $\tilde {z}$ is the quantization axis.
The small splitting of a doublet into $\Delta_a$ and $\Delta_b$ is due to
a weak in-plane anisotropy $E(S_{x}^{2}-S_{y}^{2})$.
Because of the gap anisotropy, $H_c$ depends on the
crystal orientation and ranges from about 4 T to 6 T. \cite{Honda98,Honda01,Chen}
Above $H_c$, the ordered states are commensurate with the crystal lattice for all four
magnetic-field directions studied by neutron diffraction. \cite{Chen,Zhel04}
For field $H$ applied along the $a$ axis and along $[1\bar{1}0]$,
the order is short-ranged and confined within each $bc$ plane,
whereas it is long-ranged and three-dimensional for $H||c$ and $H||[0\bar{1}1]$.

The neutron experiments have detected no incommensurate modulation
of the magnetization component parallel to the field.
Such a modulation has been predicted for the Tomonaga-Luttinger spin
liquid, \cite{Affleck91, Sakai} and its absence in NDMAP can be understood
as a consequence of a lack of axial symmetry, which is necessary for the existence
of the Tomonaga-Luttinger spin liquid. \cite{Sachdev}
It is important to note that even $H||c$ is not an axisymmetric field configuration
for NDMAP because of its unique geometry.
In this material, the crystal-field anisotropy of Ni$^{2+}$ is determined by the local
symmetry of the NiN$_6$ octahedron.
First, there is a weak in-plane anisotropy as described above.
Second and probably more important, the principal axis of the octahedron is tilted from
the $c$ axis by 15$^\circ$ toward the $a$ axis,
with the tilt direction alternating from chain to chain. \cite{Monfort}
Therefore, there is no magnetic-field direction that strictly satisfies axial symmetry.

In this paper, we explore the field-induced LRO states to 32 T
in field and 150 mK in temperature by means of specific heat measurements.
The present study greatly extends the $H$-$T$ parameter space for the
phase diagram, which has been previously limited  \cite{Honda98,Honda01,Chen}
to 12~T in $H$ and 0.52~K in temperature $T$ except for a few isolated
points obtained by neutron diffraction \cite{Chen,Zhel04}
at lower temperatures with less precision.
Preliminary results have been reported in Ref.~\onlinecite{Tsujii03}.

\section{\label{sec:2}Experimental Procedures and Results}

The single crystals of NDMAP used in this work were grown from aqueous
solutions by the method
described in Ref.~\onlinecite{Monfort}. Fully deuterated crystals
were used to eliminate the nuclear specific heat of protons
in the set of measurements extending to the lowest temperature,
whereas a hydrogenous sample was used when adequate.
The specific heat measurements to 18 T were performed in a superconducting
magnet at temperatures down to 150 mK with a dilution refrigerator.
The relaxation calorimeter
for this setup has been described in Ref.~\onlinecite{Tsujii03B}.
For measurements
from 20 T to 32 T, another relaxation calorimeter with a built-in $^3$He
refrigerator was used in a resistive Bitter magnet.
Each sample for the $H||c$ configuration was attached with silver paint \cite{Arzerite}
to the vertical face of a small silver bracket, whose horizontal face was in turn
glued with a Wakefield compound \cite{Wakefield} to the calorimeter platform.
The sample for the $H||a$ configuration was mounted with silver paint on
a piece of 0.13~mm-thick sapphire substrate, which in turn was glued with a Wakefield
compound to the platform.
A total of three samples ranging in mass from 4 mg to 8 mg
were used to cover the different field ranges and orientations.

\begin{figure}[btp]
\begin{center}\leavevmode
\includegraphics[width=0.8\linewidth]{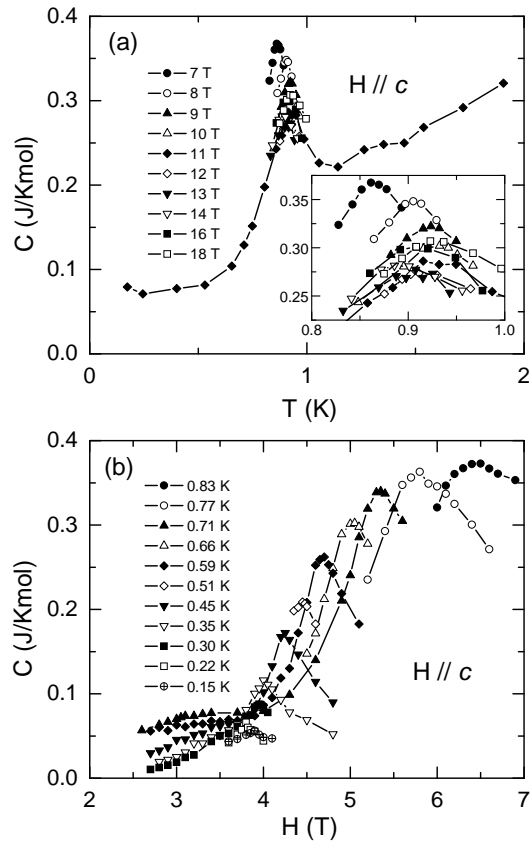}
\caption{ Specific heat of deuterated NDMAP for a magnetic
field parallel to the $c$ axis. (a) Temperature
dependence at constant fields. The inset gives an expanded
view of the region near the specific heat peak.
(b) Magnetic-field dependence at constant temperatures.
The lines are guides for the eye.
}\label{fig1}\end{center}\end{figure}

The magnetic-field and temperature dependence of the specific heat of a deuterated
sample is shown in Fig.~1 for fields up to 18 T, applied along the $c$ axis of the crystal.
At each field, the peak in the specific heat seen in Fig.~1(a) clearly indicates a phase
transition.
As can be seen in the inset, the transition temperature denoted by the peak position first increases with
increasing field up to about 10 T and then decreases for fields up to 14 T, where it starts
to increase again, making a shallow local minimum at around 14 T.
Furthermore, the peak height has a less pronounced minimum at roughly the same field.
These features reveal the existence of a new phase boundary, which separates two
field-induced ordered states at around 14 T.
For temperatures less than 0.85 K, specific heat is shown as a function of the magnetic field
in Fig.~1(b).
Again, the peak clearly indicates a phase transition.
We have confirmed that the transition temperatures determined by such magnetic-field
scans agree with those determined by the temperature scans.

\begin{figure}[btp]
\begin{center}\leavevmode
\includegraphics[width=0.8\linewidth]{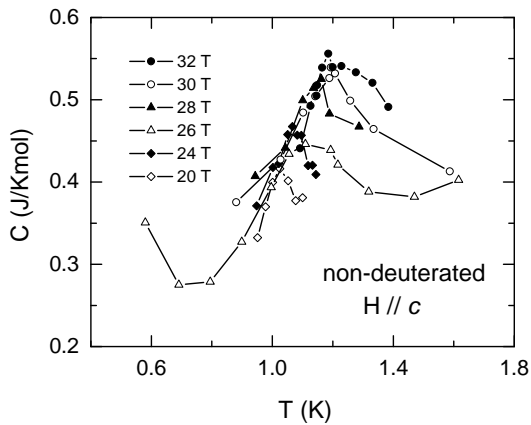}
\caption{ Specific heat of hydrogenous
NDMAP at constant magnetic fields ranging from 20 T to 32 T,
applied along the $c$ axis. The lines are guides for the eye.
}\label{fig2}\end{center}\end{figure}

The specific heat for the same field direction at higher fields produced by
the resistive magnet are shown in Fig.~2.
These data were obtained with the hydrogenous sample.
At 26 T, the nuclear contribution of the protons is visible at temperatures below 0.7 K.
However, the peak due to transition stands out, since it occurs at a higher temperature.
In this field region extending from 20 T to 32 T, the transition temperature obtained from
the peak position varies only monotonically, and so does the peak height.

\begin{figure}[btp]
\begin{center}\leavevmode
\includegraphics[width=0.7\linewidth]{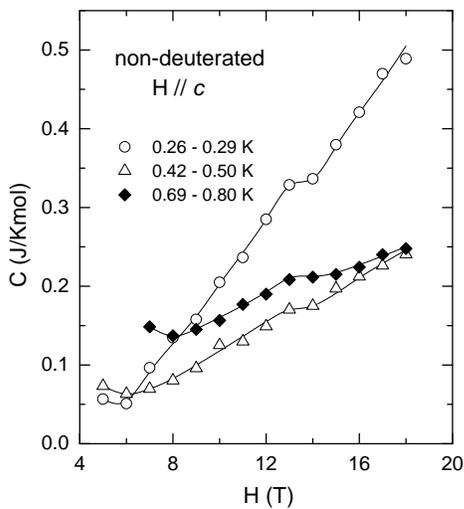}
\caption{ Magnetic-field dependence of the specific heat of hydrogenous NDMAP in the
ordered phases.  The field was applied along the $c$ axis.
The anomaly at the lowest field of each curve is due to
the proximity of the transition to the disordered spin-liquid phase.  The uninteresting proton
contribution raises the curve at 0.26--0.29 K with respect to
those at higher temperatures. The lines are guides for the eye.
}\label{fig3}\end{center}\end{figure}

To further investigate the phase diagram near 14 T,
we have measured the specific heat of the same hydrogenous sample
as a function of the magnetic field, again applied along the $c$ axis,
but to higher fields than in Fig. 1(b).
In these measurements, a constant electric current was fed to the heater of the thermal
reservoir of the calorimeter, allowing the temperature to rise monotonically with an increasing
field as dictated by the magnetoresistance of the heater.
As seen in Fig.~3, a plateau-like anomaly
in the specific heat occurs at around 14 T, clearly indicating a phase transition.
This anomaly was overlooked in our preliminary report, \cite{Tsujii03} where
the field range investigated was too narrow.
Within the experimental resolution, no corresponding feature is found
in magnetization, which has been measured at 80~mK and 1.3~K
using a pulsed magnet and is featureless up to 60 T except for an
anomaly at $H_c$. \cite{Honda00M}

\begin{figure}[btp]
\begin{center}\leavevmode
\includegraphics[width=0.8\linewidth]{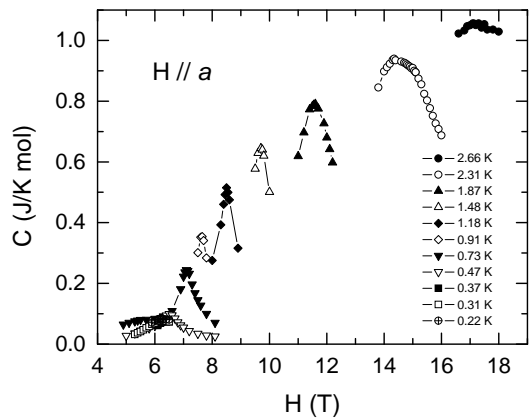}
\caption{ Magnetic-field dependence of the specific heat of deuterated NDMAP
measured at constant temperatures. The magnetic field was
applied along the $a$ axis. The lines are guides for the
eye.}\label{fig4}\end{center}\end{figure}

For magnetic fields applied along the crystallographic $a$ axis, which is perpendicular to the
spin chains, another deuterated sample was used to measure the specific heat
at temperatures below 2.7 K as a function of field, as shown in Fig. 4.
Again, a transition to the field-induced ordered phase is clearly indicated
by a peak at each temperature.
For fields ranging from 6.2~T to 6.5~T, these measurements were supplemented
by temperature sweeps similar to those shown in Fig. 1(a).
No new phase boundary was found for this field direction up to 18~T.

\begin{figure}[btp]
\begin{center}\leavevmode
\includegraphics[width=0.75\linewidth]{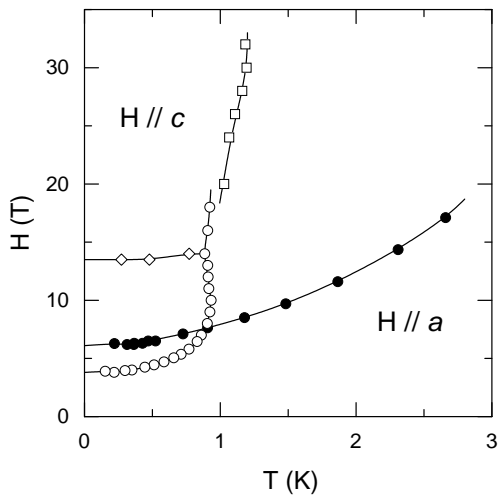}
\caption{Field-temperature phase diagram of NDMAP.
The open symbols are for the field applied along the $c$ axis and
the closed symbols for the field along the $a$ axis.  The
circles are for the deuterated samples, and the squares and diamonds for
the hydrogenous sample.
The lines are guides for the eye.
}\label{fig5}\end{center}\end{figure}

The magnetic phase diagram determined from the positions of the specific-heat peaks
and the plateau-like anomalies
is shown in Fig.~5 for the two magnetic-field directions studied.
When the field is along the $c$ axis, the transition temperature exhibits a
shallow but distinct local minimum at about $H_{c2}\approx 14$~T, and
a new phase boundary extends nearly horizontally from
this minimum.
The small break in the phase boundary
between the new high-field phase and the thermally disordered phase at around 20~T
indicates that the hydrogenous
sample has a slightly higher transition temperature than the deuterated sample.
As stated earlier, the calorimeters used for the two samples were different.
However, we have confirmed in another experiment the consistency of the temperature
scales of the two calorimeters.
Therefore, the discontinuity in
the phase boundary is not an artifact and
indicates that deuterated NDMAP in fact has a somewhat weaker exchange.
When the field is along the $a$ axis, the transition temperature rises
rapidly with increasing field, with no indication of a new phase.

\section{\label{sec:3}Discussion}

This is the first time more than one field-induced ordered phase has been found
in a quasi-1D antiferromagnet for a given magnetic-field orientation.
We believe it is significant that the field direction required for the new phase is
along the crystalline $c$ axis.
This direction, parallel to the spin chains, is quite unlike the $a$, $b$,
and $[0\bar{1}1]$ directions in that the ordering field rises very rapidly
with temperature in the overall $H$-$T$ phase diagram. \cite{Honda98,Honda01,Chen}
This feature probably exemplifies the underlying tendency of the spin chains in
this field configuration to form
a Tomonaga-Luttinger spin liquid, as has been presumed earlier. \cite{Honda01}
Hence it is likely that identifying the new phase above $H_{c2}$ and understanding
the mechanism of the transition will shed light on the physics
of the Tomonaga-Luttinger spin liquid.

At present, however, with the absence of experimental
data in the field region near and above $H_{c2}$ other than the specific heat and
magnetization, which shows no anomaly
at $H_{c2}$, we can at best speculate on the nature of the transition.
One possible scenario can be a transition involving spin rotation around
the direction of the applied field, which is parallel to the spin chains.
Such an exotic transition has been observed for instance in a spin-1/2 chain
material BaCu$_2$Si$_2$O$_7$,
where it is presumably driven by a competition between an off-diagonal exchange and
the magnetic field. \cite{BCSO}
Although the Dzyaloshinskii-Moriya interaction, \cite{DM} the likely candidate for
the off-diagonal exchange, is usually quite
small for Ni$^{2+}$ and has not been detected in NDMAP, a spin-rotation transition
is an interesting possibility.

Recently, Wang \cite{Wang} has considered a
spin-1 chain with broken symmetry using a fermionic field theory
and has predicted that a magnetic field
larger than $H_c$ will restore an approximate axial symmetry
and lead to a well-defined second transition from a commensurate
phase to an incommensurate phase.
Unlike in the Tomonaga-Luttinger
spin liquid, the excitations in the incommensurate phase will be gapped.
However, the gap will be small and roughly equal to $(\Delta_b-\Delta_a)/2$,
when the field is applied along the chain direction $c$.
Although Wang has predicted an absence of anomaly in specific heat
--- as well as magnetization --- at the second transition, which is other than
first-order, it is worth exploring the possibility of the phase above $H_{c2}$ being
an incommensurate phase.

\begin{figure}[btp]
\begin{center}\leavevmode
\includegraphics[width=0.9\linewidth]{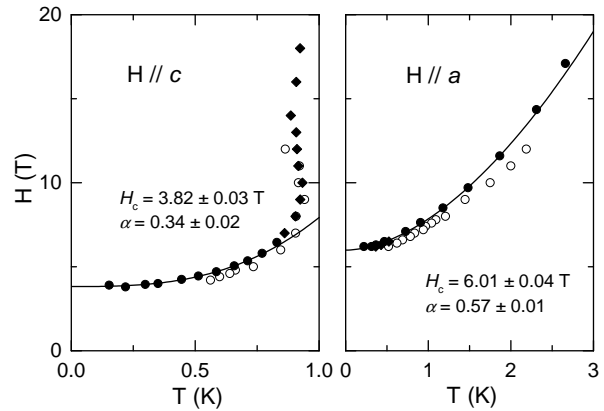}
\caption{ Field-temperature phase diagram of NDMAP below 20 T for the
field applied parallel to the $c$ and $a$ axes.
The closed symbols are for the deuterated samples of the present work, where the circles are from the
temperature-scan data and the diamonds from the field-scan data.
The open circles are for hydrogenous samples from Ref.~\onlinecite{Honda98}.
The lines are low-temperature fits of the phase boundaries of the deuterated samples to the expression
$T_c \propto (H - H_c)^\alpha$.
}\label{fig6}\end{center}\end{figure}

The Bose-Einstein condensation of triplets is believed to be a valid description of
field-induced LRO in all gapped antiferromagnets, regardless of the type of disordered
ground state below $H_c$ and the type of the ordered state above it, provided that the
lowest excitations are bosons.
According to the Bose-Einstein condensation theory, \cite{Tsvelik,Nikuni} the phase
boundary in the $H$-$T$ parameter space obeys a power law
$T_c \propto (H - H_c)^\alpha$.
The field-induced ordering in the $S=1/2$
spin-dimer material TlCuCl$_3$ was the first case that was analyzed in terms of
the Bose-Einstein condensation. \cite{Nikuni}
However, the exponent $\alpha$ of 0.50 determined \cite{Tanaka04}
for this material is smaller than the
theoretical value of 2/3 obtained by a Hartree-Fock approximation for a quadratic
dispersion for the triplets.
The exponent for
Ni(C$_{5}$H$_{14}$N$_{2}$)$_{2}$N$_{3}$(ClO$_{4}$) (NDMAZ), a
spin-1 chain material \cite{Honda97} similar to NDMAP, has been
reported to be 0.45 for the magnetic field applied along the $c$ axis.
\cite{Kobayashi}  This value is also significantly smaller than the
theoretical one.

The four-fold improvement in the minimum temperature over the previous work \cite{Honda98,Honda01}
for the field directions along the $c$ and $a$ axes allows us to determine the
field dependence of the ordering temperature near $H_c$ with high accuracy.
The relevant portions of the phase boundaries from Fig.~5 are reproduced
in Fig.~6, separately for the two field directions.  These boundaries, shown with
closed symbols, are for deuterated samples.
Excellent fits of the data below 0.77~K for $H||c$ and below 1.9~K for $H||a$
are obtained with the power law, giving $\alpha=0.34\pm0.02$ and $0.57\pm0.01$, respectively.
The critical fields are $H_c=3.82\pm0.03$~T and $6.01\pm0.04$~T, respectively,
along the $c$ and $a$ axes.

Recently, specific heat measurements of the $S = 1$ bond-alternating-chain
antiferromagnet Ni(C$_{9}$H$_{24}$N$_{4}$)NO$_{2}$(ClO$_{4}$) (NTENP)
have found $\alpha=0.334$ and 0.52, respectively, for the field parallel
and perpendicular to the spin chains. \cite{Tateiwa}
The similarity of these exponents to those for NDMAP
suggests that NTENP has similar field-induced
ordered phases, although its low-field ground state is
a spin-dimer singlet\cite{Hagiwara01} rather than a Haldane spin liquid.

Including ours, all the experimental values found to date for the exponent $\alpha$
are smaller than 2/3 predicted by the theory.
However, the values for NDMAP and NTENP for the magnetic field applied
along the chain direction agree, within the combined uncertainties of the experiments
and the calculation, with the quantum Monte-Carlo results
by Wessel~$et\ al$., \cite{Haas} who have found $\alpha = 0.37\pm0.03$
for $S=1/2$ spin dimers with a weak three-dimensional inter-dimer coupling
and $\alpha = 0.32\pm0.03$ for two-leg spin-1/2 ladders with a weak
three-dimensional inter-ladder coupling.
Their results support the Bose-Einstein condensation theory, according to the
quantum Monte-Carlo calculation for three-dimensionally
coupled $S=1/2$ spin dimers by Nohadani~$et\ al$., \cite{Nohadani} who have shown
that $\alpha$ deviates downward from 2/3, as the temperature
range of the power-law fit is widened.
Furthermore, Misguich and Oshikawa \cite{Misguich} have used a realistic dispersion for
the triplets in TlCuCl$_3$ in their calculation of the field dependence of the Bose-Einstein condensation
temperature $T_c$ and found good agreement with the experiment. \cite{Nikuni}
It remains to be seen, however, whether the Bose-Einstein condensation theory can explain
the strongly anisotropic $\alpha$ for NDMAP and NTENP, particularly the extremely small $\alpha$
when the field is applied along the spin chains.

In Fig. 6, we have included the phase boundaries for hydrogenous samples from
the previous work, as shown with open symbols. \cite{Honda98,Honda01} Again, the ordering temperature
is somewhat higher for the hydrogenous samples except at the fields of 10 T and 12 T
applied parallel to the $c$ axis, whereas $H_c$ is identical for a given field direction
within our accuracy.
This indicates that the interchain exchange $J'$ is slightly weaker in deuterated NDMAP than in
hydrogenous NDMAP, since decreasing $J'$ lowers the ordering temperature
directly but has little effect on $H_c$, which is primarily determined by
the in-chain exchange $J$.
A previous study, \cite{HondaUnpub} which compared the phase diagrams
of deuterated and hydrogenous samples for $H$ along the $b$ axis, saw
a much smaller but qualitatively similar difference.

We propose that the smaller $J'$ for deuterated NDMAP is caused by
the smaller zero-point motion of deuterons leading to less overlap of
the electronic wavefunctions in the exchange paths involving hydrogens.
In some classical antiferromagnets, such as the linear-chain compound
(CH$_3$)$_4$NMnCl$_3$ (TMMC), deuteration has been reported to reduce the
N$\acute{\rm e}$el temperature. \cite{Birgeneau,Boucher}
In the quasi-two-dimensional organic salt $\kappa$-(BEDT-TTF)$_2$Cu[N(CN)$_2$]Br,
substituting deuterium for the hydrogen of the ethylene groups has been shown to drive the system
from a superconductor to an antiferromagnetic Mott insulator. \cite{Taniguchi}
To our knowledge, however, the present
work represents the first observation of an effect of
deuteration in a Haldane-gap antiferromagnet.

In summary, the magnetic phase diagram of NDMAP has been
extended by specific-heat measurements to 150~mK in temperature and
32 T in the magnetic field.  A new transition has been found at
$H_{c2}\approx 14$~T, when the field is parallel to the spin chains.
Further study using techniques other than
specific heat is needed to investigate the nature of the new phase,
including the possibility that it is an incommensurate phase.
In addition, we have determined the power-law exponents for the transition
temperatures of the field-induced ordering from the spin-liquid state
and compared them with the exponents for TlCuCl$_3$, NDMAZ, and NTENP
and with the predictions of the Bose-Einstein condensation theory.
Finally, we have observed an effect of deuterium substitution on the
ordering temperature.  The result indicates that the
substitution slightly decreases the interchain exchange $J'$ but hardly
affects the in-chain exchange $J$.

\begin{acknowledgments}
We thank S.~Haas, M.~Hagiwara, I.~Harada, M.~W.~Meisel, D.~A.~Micha, Y.~Narumi,
J.~Ribas, T.~Sakai, D.~R.~Talham, and H.~Tanaka for useful discussions, 
and F.~C.~McDonald,~Jr., T.~P.~Murphy, E.~C.~Palm, and C.~R.~Rotundu for assistance.
This work was supported by the NSF through DMR-9802050 and the DOE under Grant No.
DE-FG02-99ER45748.
A portion of it was performed at the National High Magnetic
Field Laboratory, which is supported by NSF Cooperative Agreement
No. DMR-0084173 and by the State of Florida.
KK acknowledges support by a Grant-in-Aid for Scientific Research
from the Japan Society for the Promotion of Science.
\end{acknowledgments}

\end{document}